# TCP-UB: A New Congestion Aware Transmission Control Protocol Variant

[1]Wafa Elmannai, [2]Abdul Razaque, [3]Khaled Elleithy

Department of Computer Science and Engineering, University of Bridgeport,
Bridgeport, CT 06604, USA
[1]welmanni@bridgeport.edu, [2]arazaque@bridgeport.edu,
[3]elleithy@bridgeport.edu

## ABSTRACT

*Transmission control protocol (TCP) is a connection oriented protocol for several types of distributed applications. TCP is reliable particularly for traditional fixed networks. With emergence of faster wireless networks, TCP has been performing poorly in its original format. The performance of TCP is affected due to assorted factors including congestion window, maximum packet size, retry limit, recovery mechanism, backup mechanism and mobility. To overcome deficiency of original TCP, Several modifications have been introduced to improve network quality. The mobility is a major hurdle in degrading the performance of mobile wireless networks. In this paper, we introduce and implement new TCP variant University of Bridgeport (UB) that combines the features of TCP Westwood and Vegas. We examine the performance of TCP-UB, Vegas and Westwood using different realistic scenarios. NS2 simulator demonstrates the stability of TCP-UB as compared with TCP Vegas and Westwood in highly congested networks from the mobility point of view.*

## KEYWORDS

*TCP UB, TCP Westwood, TCP Vegas, mobility, Bandwidth estimation.*

## 1. INTRODUCTION

Deployment of emerging wireless network technologies has motivated many researchers to introduce a new transport protocols for faster communication. The original TCP was particularly designed for fixed wired networks [8]. TCP in its current shape is not an optimal transport service provider for mobile and wireless networks. Several TCP variants have been proposed and implemented in order to augment the performance [4].

In 1988, TCP Tahoe was the first version introduced by Jacobson [3]. Later other variants have been implemented such as: Reno, New Reno and Sack. In 1994, Brakmo implemented TCP Vegas [6]. The most improvement that could make the Vegas variant more powerful in that time is the addition of new congestion avoidance mechanism [13]. With the usage of this mechanism, Vegas work in completely different way from other variants.

TCP Vegas is a promising featured variant that increases the throughput performance and reduces packet loss. The advantage of TCP Vegas is to calculate the available bandwidth in network. TCP Vegas determines the bandwidth on the basis of difference between expected and actual throughput to avoid packet loss [6].

TCP Westwood is another variant that gives more significant improvement in wireless networks with lost links [1]. Westwood is the sender side modification of TCP Reno. The beauty of TCP Westwood is to use bandwidth estimation at the sender side. It defines as bottleneck sharing the connection of the network to coordinate both slow start threshold (ssthresh) and congestion window size (CWND). The estimation of Westwood is based on measured acknowledgment (ACK) [7].

Whenever, the congestion occurs in the network, TCP Westwood resets both (ssthreshold) and (CWND) based on the bandwidth estimation. When ACK is received, the sender provides information about the transmitted data at the receiver side. If no loss of packets is notified that shows fairness for bandwidth estimation. If duplicated ACK is received, new bandwidth estimation should be reset after resending the lost packet. But it is difficult for sender to determine which packet causes of duplicated ACK.

It has been observed that TCP Vegas performed better than Reno with 4.29% to 9.7%, SACK with 1.64% to 4.66%, Tahoe with 4.11% to 9.71% and Westwood with 1.12% to 2.9% & New Reno with 2.01% to 5.6%. [8]. However, the minimum effect of mobility has been recorded on TCP Westwood. So, the scope of our research is to introduce a new TCP variant to provide better performance based on the above results that we obtained.

In this paper, we propose new TCP variant integrating the important features of TCP Vegas and Westwood to gain better performance from both efficiency and mobility point of view. The remaining paper is organized as: the related work will be discussed in section 2, the proposed work is in section 3, over view of mobility model and simulation setup are explained in section 4, the simulation results and analysis are given in 5 and finally section 6 concludes the paper.

## 2. RELATED WORK

In [9] the performance of congestion control algorithms were examined for both TCP Vegas and TCP Bic by using ns2 over mobile adhoc networks (MANETs). The major focus of the paper was to determine the loss of packets, round trip time (RTT), measuring the control window (CWND) and throughput. TCP Vegas proved a high performance during a whole period of simulation as compared with TCP Bic. TCP Bic improved the performance after half of simulation time but it did not compete with TCP Vegas.

Multiple Paths TCP Westwood (MPTCPW) was proposed in [10] to improve the performance of TCP Westwood over wireless networks. Furthermore, the proposed scheme controls the congestion over different paths to solve the bottleneck problem as regular Westwood cannot solve this problem. The proposed work introduced new features in TCP Westwood such as: load-balancing and fair sharing at bottleneck. After introduction of new features, the scheme compared both multiple paths TCP (MPTCP) and MPTCPW. Thus, MPTCPW could increase the throughput and attain the stability over number of conditions.

Combined rate and bandwidth estimation (CRB) were proposed in [14]. This method examined whether the acknowledgments spread out evenly over the time or not, then used the technique to detect the error. This proposition helped to modify the TCP Westwood in order to increase efficiency and friendless with respect to TCP Reno adjustment. Therefore, the results could prove a good performance for different types of errors and control the adjustment of friendless and efficiency.

A new algorithm TCP Westwood (TCPW-M) was proposed in [12]. It improved the performance of regular TCP Westwood over the hybrid networks. TCPW-M could improve the bandwidth estimation and increase the throughput of Westwood. On basis of simulation analysis, TCPW-M achieved higher stability over the level of different measuring conditions.

The performance of TCP Vegas in two scenarios was analyzed and exanimated over large delay network and large bandwidth using ns2 simulator [2]. The experiment used various values of the based parameters (Alpha and Beta) to measure the congestion window in both phases, slow start and congestion phase. The simulation results proved that TCP Vegas with default parameters could not achieve high performance with large bandwidth. When they increased the values of Alpha and Beta, TCP Vegas could attain higher performance.

TCP Vegas was also compared with TCP-Reno, TCP-New Reno and TCP Sack over Mobile Ad hoc Networks (MANETs) in [11]. The experiment analyzed and studied congestion avoidance and congestion control mechanisms. However, the results showed that TCP Vegas achieved higher performance over the comparison. It provided less packets loss with faster loss detection as compared with other TCP variants. Therefore, we proposed new efficient TCP-UB variant that combines the features of both TCP Westwood and TCP Vegas.

## 3. PROPOSED WORK

### 3.1. Description of Hybrid Network Design

Our objective in this research is to introduce a new mechanism achieve faster and reliable communication in MANET areas. This result will allow users to deploy MANET in the educational field to realize robust and reliable delivery of data.

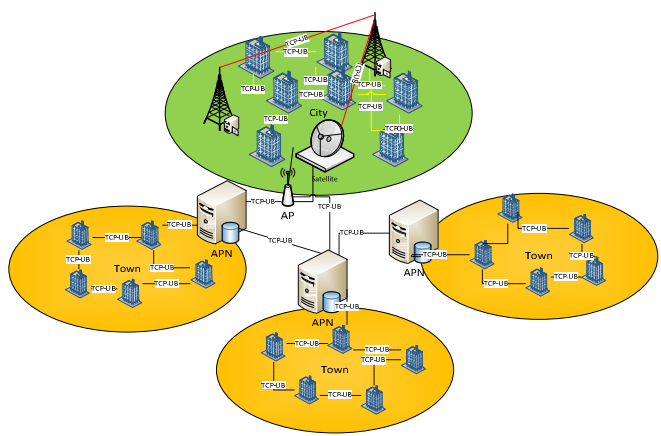

Figure1: Hybrid network design

We design hybrid network that includes wireless and MANET networks as shown in figure1, in order to make a faster communication between the nodes in the MANET area. Our research objective is to provide a communication for scatted educational institutions. Our designed network is based on one city that represents the wireless network to provide the communication and numbers of towns are displayed as MANET area.

We have used Anchor Point Node (APN) [8] to make communication between all MANET and wireless network as well as between the MANET networks themselves. An APN can be located in different places based on the number of networks inside the hybrid network. Hence, APN can be considered as component of MANET area which called MANET Anchor Point Node (MAPN) and for wireless network as Infrastructure Based Anchor Point Node (IBAPN). APN can get the nodes information since each node gets its own IP from Dynamic Host Configuration Protocol (DHCP). Based on this information TCP-UB can provide a stable connection between the nodes with each other inside the MANET area (Town).

Furthermore, MANET nodes are mobility aware but even though they do not lose the data since TCP-UB controls the communication between the nodes. Hence, TCP-UB proves less mobility effect which reduces affection of the mobility on the routing. It can make the percentage of delivering the packet higher with less loss. When the TCP-UB coordinates the communication between the nodes in the MANET areas, APN controls the connection between the wireless and the MANET networks that increases the efficiency of network. With respect to the mobility point, we have used RW model inside the MANET areas to control the speed and the mobility of the nodes. We have also deployed constant-bit rate (CBR) to route in MANET areas. This network supports for several educational institutions for sending and receiving data.

### 3.2. TCP-UB algorithm

The behavior of TCP-UB is based on three partitions of congestion phases shown in Figure2. Each part is considered as threshold to initiate separate process for preventing the congestion. In our case, the process of treating congestion window at three separate phases saves the bandwidth.

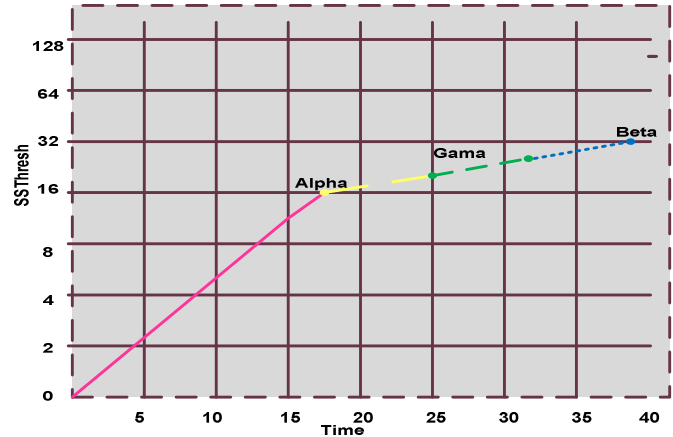

Figur2: The Congestion Control Behavior of TCP-UB

First, the process of TCP-UB starts with computation of initial congestion on basis of CWND and the Base RTT.

Expected Rate= CWND /Base RTT ------------------------------------------------------------------- (1)

Where, the expected rate means predicted data rate to be sent on the congestion window based on previous condition of network. Expected rate is obtained with congestion window and Base RTT. The congestion window size (CWND) means the size of the window when the data is transferred over the time and Base RTT is minimum RTT which required for updating the congestion window size.

When we obtain the expected rate of the congestion window, the next step is to find out the rate of current Window (actual rate) which means the current data rate. That is sent on the current window over the network. The actual rate is obtained by dividing the CWND by actual RTT. When the actual RTT is the time that takes to transfer the packet and receive the acknowledgment in current window.

Actual Rate= CWND/RTT ------------------------------------------------------------------------------ (2)

Then we determine the difference (Diff) between the actual and expected rate and multiplying Base RTT.

Diff = (Expected Rate – Actual Rate) BaseRTT ------------------------------------------------------- (3)

The window size is updated based on (3) by checking the difference with each one of the three thresholds (Alpha, Beta, Gama). First, check if the difference is less than the minimum threshold (Alpha).

If (Diff < α) then-------------------------------------------------------------------------------------- (4)

If step number 4 is met then increase the CWND by one.

CWND+1 ------------------------------------------------------------------------------------------------- (5)

132

If not so, compare the difference with the middle threshold (Gama) if they are equal or not. Where, Gama is used to estimate the congestion possibility.

If (Diff=$\delta$) then -------------------------------------------------------------------------------- (6)

Based on (6), if it is met, we let the slow start threshold (ssthresh) equal to bandwidth estimation which is the sharing of the bottleneck that used by the connection of the network, multiplying the size of segment which means the size of the packet that will be sent . The usage of the ssthresh is to find out whether TCP-UB is in slow start phase or congestion phase.

Let ssthresh = (BWE*Base RTT)/ seg_size -------------------------------------------------------- (7)

Then assign the value of ssthresh to CWND if the CWND is bigger than ssthresh.

If (CWND > sthresh) then

CWND=ssthresh --------------------------------------------------------------------------------- (8)

If the difference is not equal to Gama then there is possibility of congestion and TCP-UB needs to check the time out if it is expired or not. The time out is the time limit of transmitting the packet and receiving the acknowledgment within this time unless that we need to use retransmission mechanism.

If (the time out is expired) then ------------------------------------------------------------------ (9)

 If it is expired then assign value one to CWND.

Let CWND=1-------------------------------------------------------------------------------------- (10)

Let's ssthresh equal to the bandwidth estimation multiplying Base RTT then divide that over the size of the segment.

ssthresh = (BWE*BaseRTT)/seg_size ----------------------------------------------------------- (11)

And assign value two to the ssthresh if the ssthresh is less than two.

If (ssthresh<2) then

Ssthresh=2 ------------------------------------------------------------------------------------- (12)

Third comparison is between the difference and the maximum threshold (Beta).

If (Diff >$\beta$) then ------------------------------------------------------------------------------- (13)

If the difference is bigger than maximum threshold (Beta) decrease the CWND by one.

CWND-1------------------------------------------------------------------------------------------ (14)

Finally, if all the three conditions are not applied then CWND should be fixed.

Otherwise -> CWND ---------------------------------------------------------------------------- (15)

## 3.3. Bandwidth Estimation:

The bandwidth estimation process is used for measuring the acknowledged and transmitted packets. Our goal is to measure the bandwidth for transmitted packets. The following algorithms help to update the congestion window, slow start threshold and determining the bandwidth.

**Algorithm1**: Updating congestion window and
slow start threshold for transmitted packets
1. If (Packet is sent)
2. sample_length[M] = (packet_size*8);
3. sample_interval[M] = now - last_sending_time;
4. Average_packet_length[M]= alpha*
   Average_packet_length[M-1]+(1-alpha)
   *sample_length[M];
5. Average_interval[M] = alpha* Average_interval
   [M - 1]+ (1-alpha ) * sample_interval[M];
6. Bwidth[M] = Average_packet_lengthM]/
   Average_interval [M]
7. Endif
8. sample_length[M] = (acked * packet_size * 8);
9. sample_interval[M] = now - last_ack_time;

Where, last_ack_time = Time for last received acknowledgement. Acked = number of
acknowledged segments when congestion occurs. On the basis of updating, the cwnd and
ssthresh are given in algorithm 1. Bandwidth estimation process is determined with following
algorithm.

**Algorithm 2**: Estimating bandwidth of
transmitted packets

1. f (3 duplicate ACKs are received)
2. ssthresh = Bwe * RTT_min
3. if (cwnd > ssthresh)
4. cwnd = ssthresh
5. end if
6. end if
7. if (retransmission timeout expires)
8. ssthresh = Bwe * RTT_min
9. cwnd = 1
10. end if

## 4. OVER VIEW OF MOBILITY MODEL AND SIMULATION SETUP

The idea of this paper is to introduce a new TCP Variant over a hybrid network. In this paper,
we have compared the performance of our new TCP-UB protocol with Vegas and Westwood
from mobility point of view. We mainly focus to analyze the performance and efficiency of
each variant using different type of scenarios. To validate the performance, we measure
interesting parameters include goodput, effect of mobility, congestion control, and bandwidth
consumption.

### 4.1. Overview of Random Waypoint model (RW)

Several mobility models have been discussed in [5] including manhattan mobility model (MM),
free mobility model (FM) and reference point group mobility model (RPGM). But (RW) is one
of the widely used mobility model in adhoc mobile network (MANET). It is introduced by
Johnson and Maltz. The beauty of this model is to control the movement of the nodes over different number of paths and velocity. Nodes are distributed either normal or uniform way in RW. In RW, the pause time is randomly selected; nodes can stop during that time whenever they reach their destination. This process is repeated until simulation time ends.

## 4.2. Simulation Set Up

The scenarios are simulated using ns2.28 on LINUX Red Hat-9. We generate Random Way Point Mobility Model (RW) to control the movement of the nodes. We also have implemented TCP-UB algorithm by Object-Oriented extension of TCL (OTCL) as shown in figure3. TCP-UB, TCP Vegas and TCP Westwood are simulated over the network to compare the performance of each of them over the measured conditions. The propagation and transmission rang is 250 meter. 100 nodes are placed over wireless and MANET network. The length of the packet is 1040 bytes included 40 bytes payload within square of 1000*1000 meter. The nodes cannot transmit after this limit. The minimum speed of the mobile node is 0 m/sec and the maximum speed is 35m/sec. Each node can send 8packets/sec. The simulation time has been set as 140 seconds. Random Waypoint Model (RWM) is imitated for starting nodes' location. The pause time has been set as constant value 5 seconds for each 50 seconds. By dividing both the minimum and the maximum speed of the node [Vmin, Vmax], we can get the moving speed randomly. The buffer size of the queue is 80 packets. Antenna Type is Omni directional.

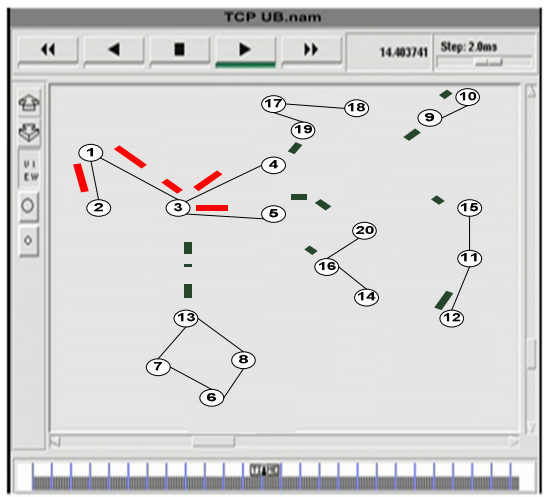

Figure3: The implementation of TCP-UB over Wireless and MANET Networks.

## 5. SIMULATION RESULTS AND ANALYSIS

In this section we discuss the simulation scenarios.

## 5.1. Efficiency Variance Scenario:

In this scenario, we have simulated our network over MANET and wireless segments using NS2, and examined the efficiency of TCP Westwood, TCP Vegas and TCP-UB. For each of above TCP variants, we have collected their acknowledged and received packets. In this scenario, the average speed is 17.5 m/sec for each TCP variants with Random Waypoint Mobility model.

Figure 4 shows the efficiency of TCP Vegas, which steadily decreases for acknowledged packets from 5.8Mb to 4.2Mb over the time. In Figure 5 we can notice that the efficiency of TCP Westwood decreases with almost the same numbers. The reason for this decreasing of packet's efficiency is the mobility. This scenario covers MANET and wireless. In this condition, MANET stays dynamic make radio channel fading and mobility of nodes are main effects.

The mobile nodes take longer time to recover from broken links. The mobility of nodes has an effect on TCP variants due to changes of routing information over the network and can cause longer RTT and repeated timeouts. In fact, the mobility of nodes can make the receiver getting out of order packets which can affect the acknowledgements. However, that can cause the duplicating acknowledgement and starting retransmission algorithm with reducing in the congestion window [8].

On the basis of efficiency, it is clear that TCP-UB acknowledges more packets than TCP Vegas and TCP Westwood as shown in Figure 6. These data shows TCP-UB received and acknowledges more packets compared with other variants. Furthermore, an important feature of TCP-UB is the stability. The performance of TCP-UB becomes stable during all the simulation time.

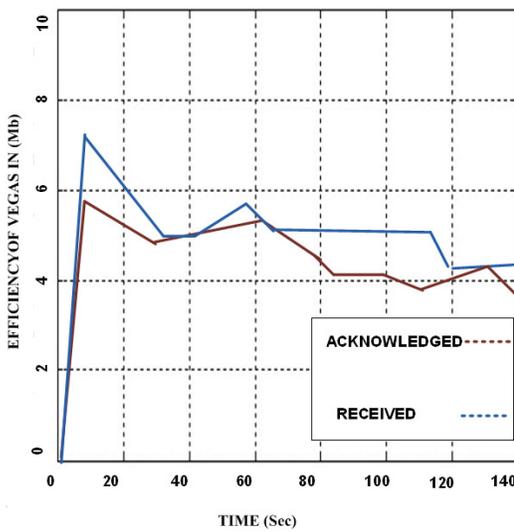
Figure 4: show the efficiency of Vegas

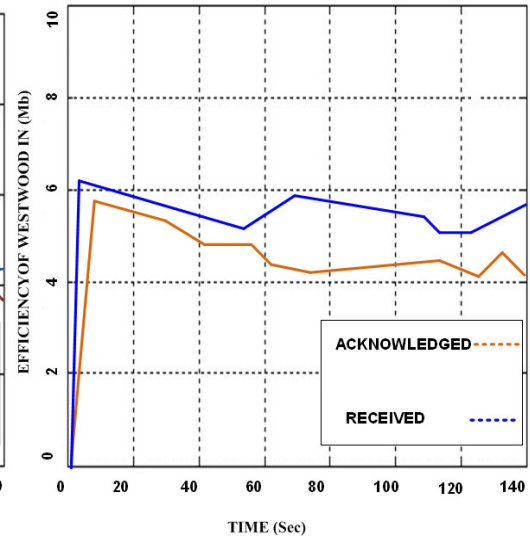
Figure 5: show the efficiency of Westwood

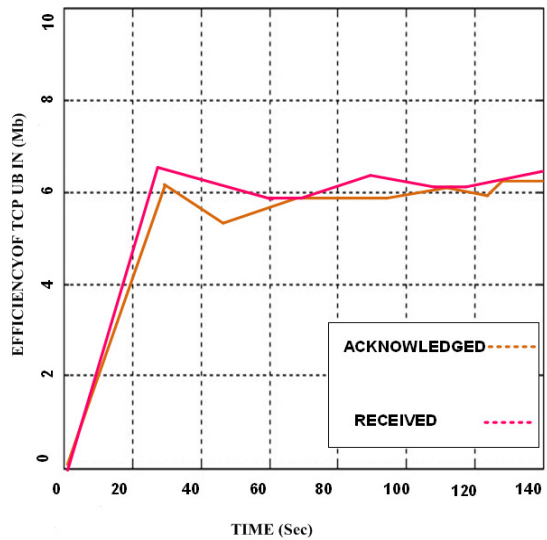
Figure6: show the efficiency of TCP-UB

## 5.2. Goodput Scenario

We show average of goodput for TCP Vegas, TCP Westwood and TCP-UB from static and mobility point of view as are shown in Figure 7 and 8.TCP Vegas and Westwood are not stable if the speed increases from 25 to 35m/sec. They show poor performance while TCP-UB the most stable performance throughout changes in nodes' speed. The changes in speed do not affect the performance of TCP-UB because including of Gama, the goodput of TCP- UB is better than other TCP variants.

Another important factor is using Gama for division of congestion avoidance phase into three parts. The partition of congestion avoidance phase provides sufficient time to control congestion window and loss of packets. The behavior of routing protocols also cannot affect Good put performance of TCP-UB. The performance of TCP-UB, TCP Westwood and TCP Vegas is shown in Figure 7 for Mobility view and in Figure 8 for static view.

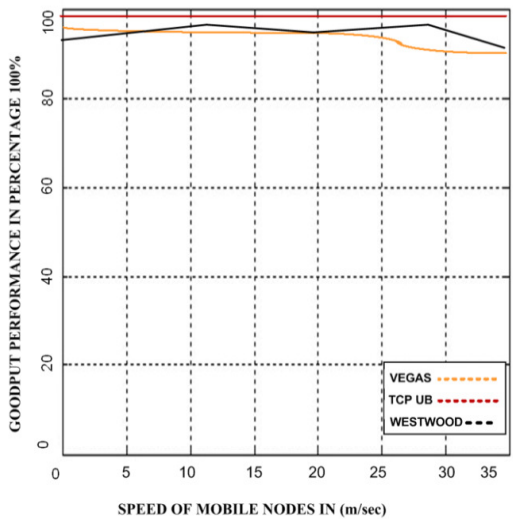
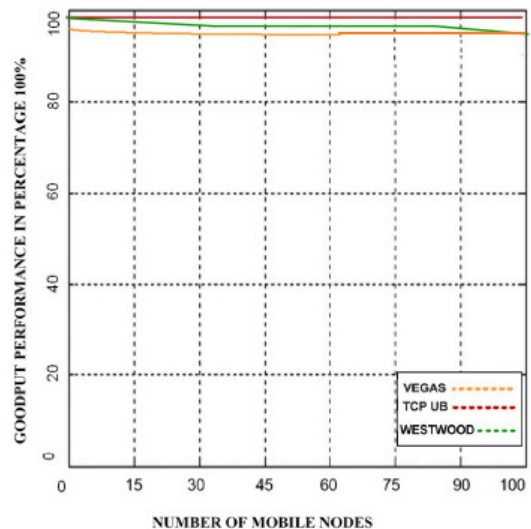

Figure 7: Mobility scenario of Goodput          Figure 8: Static Scenario of Goodput

## 5.3. The Congestion Window Scenario:

In this scenario, we have studied the congestion window algorithm for TCP-UB, Vegas and Westwood. We examined the congestion algorithm over two scenarios which are static and mobility based. On basis of several literature surveys, we have found that some of congestion control algorithms have capability to improve the congestion window but they still face some experiences. The reason of this experiment is to compare our variant with well-known TCP variants: Vegas and Westwood. This experiment gives validation about suitability of proposed and existing variants. We are able to find out how to manage the congestion in busty traffic scenarios. Based on our result for both scenarios, we validate on the basis of findings that no TCP congestion based algorithm could perform very well all the time in the hybrid network.

Figure 9 shows the mobility scenario of congestion window for TCP-UB, Vegas and Westwood. When TCP-UB achieves a good performance and maintains the congestion window as compare with Vegas and Westwood. TCP-UB stays stable over the most of time of simulation. It gives 96% fairness (from 45 to 78 seconds) and 95% (from 110 to 140 seconds). TCP Vegas stay stable in one case which gives 83% (from 60 to 90 seconds). TCP Westwood doesn't execute well over the network in both cases, which are multiple loss segments and congested links.

For static scenario, figure10 demonstrates that TCP-UB achieves the stability over the most of time during the simulation. TCP-UB attains 95% (from 10 to 70 seconds) as well as 94% (from 110 to 140 seconds). TCP Vegas and Westwood perform badly over the congested network due to mobility and bandwidth estimation issues. TCP Vegas performs worse because it gets lower share of bandwidth utilization with other variants. Therefore, it achieves degraded performance. So, it is clear to prove that TCP-UB enhances the network performance and attain higher utilization of resources over hybrid network.

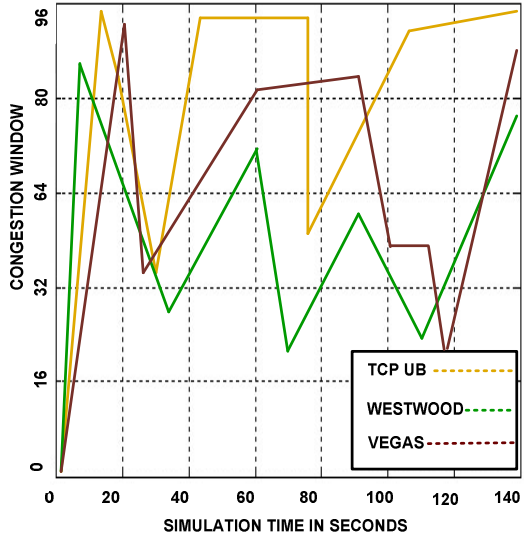 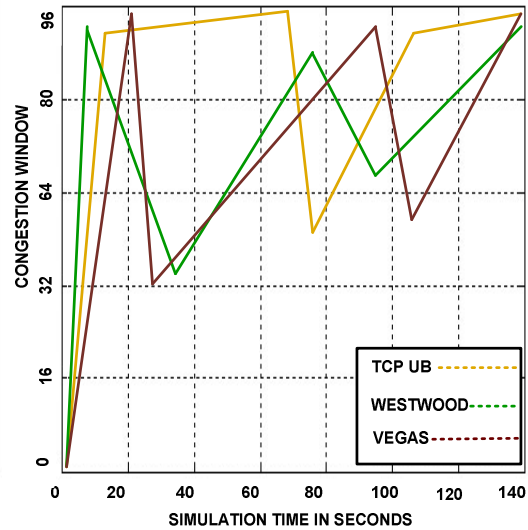

Figure 9: Mobility Scenario of CWND    Figure 10: Static Scenario of CWND

## 5.4. Bandwidth Consumption Scenario:

In this scenario we have compare the bandwidth consumption for each TCP Vegas, Westwood and TCP-UB as shown in Figure11.

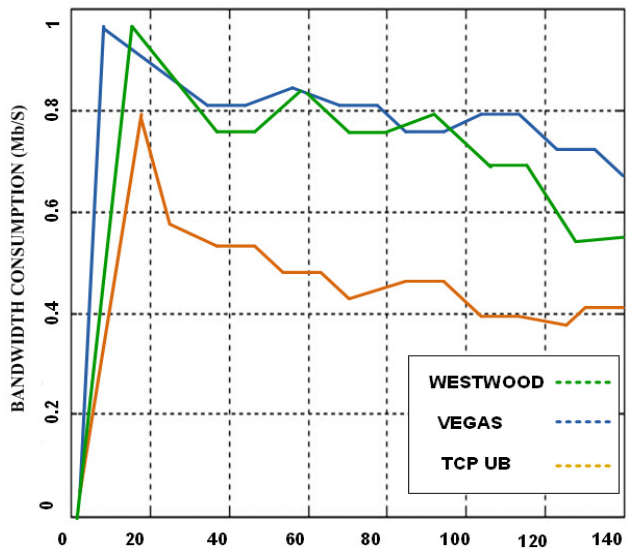

Figure 11: Bandwidth Consumption Scenario

Each of them has its own estimation algorithm. TCP Vegas has a good bandwidth estimation algorithm. This algorithm is based on the difference between the expected and actual rate. TCP Vegas change the CWND based on this difference. That means whether actual rate met the expected calculation or not. when Vegas shares the connection with other variants within one network, It starts getting aggressive which make the Vegas consumes higher bandwidth as shown in Figure11.

Also, TCP Westwood has a good bandwidth estimation algorithm since it using goof recovery mechanism. Westwood gets this estimation basis on acknowledgments ration. Estimation algorithm uses to set ssthreshold and CWND. So, whenever the congestion occurs we need to reset the bandwidth estimation. When duplicated ACK arrived, TCP Westwood always starts recovering from unacknowledged packet even if it is received by the receiver. That make the Westwood consume much bandwidth

However, TCP-UB proved less bandwidth consuming since it is integrated of both Vegas and Westwood. TCP-UB could get the good features from both variants. With division the portion after the slow start threshold, based on Gama TCP-UB can manage the movement from Alpha to Gama and from Gama to Beta. If there is congestion detected, we can reduce the CWND by one. If not so, then based on the estimation of available bandwidth CWND can be increased by one and send more packets. That can make the bandwidth consumption less. Also, in congested network, TCP-UB does not need to send all unacknowledged packets which. It will send only the lost packets since TCP-UB has the transmitted packet information. Finally, it is clear that TCP-UB consume less bandwidth by integrating Vegas and Westwood features together.

## 6. CONCLUSION AND FUTURE WORK

In this paper, we have introduced and implemented a new TCP-UB variant by amalgamating the features of TCP Vegas and TCP Westwood. We compare the performance of TCP-UB with TCP Vegas and TCP Westwood based on different simulation scenarios. Efficiency, goodput, performance, congestion control, and bandwidth consumption parameters have been tested. The results demonstrate that less effect of mobility is measured on TCP-UB. It achieves high efficiency and higher delivery of data as compared with TCP Vegas and Westwood. TCP-UB leads to a fair allocation of consumption the bandwidth. Furthermore, TCP-UB proves the stability for a longer period in the congested network for most of time. All testing scenarios demonstrate that TCP-UB is highly promising variant in adhoc networks (MANET). It is deployed in scattered educational institutions but applications of TCP-UB can be introduced for military environment, scattered hospitals and industrial zones for faster and reliable communication.

The implementation of this algorithm shows that there are still more challenges related to TCP variants that can be addressed in the future. We are planning in the future to implement TCP-UB in highly congested scenarios increasing the number of nodes especially in battlefield environments.

## AUTHORS

**Mrs. Wafa Elmannai** is Master student in the School of Engineering and Computer Science at the University of Bridgeport. She has finished her Bachelor degree in Computer Science during 2005 at Ben Ashore College and awarded high merit certificate.

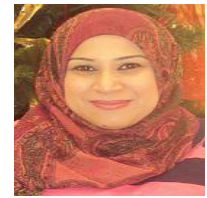

Mrs. Elmannai worked as Research Assistant at University of Bridgeport during 2010-2011. She started her career as assistant administrator in management department of education from 2005 to 2009. She also served as teacher in Algamaheria School during 2004-2005. The research interest of Mrs. Elmannai includes transmission control protocols (TCPs), mobile wireless communication, and design of mobile applications. She has published several papers in international conferences and Journals and made presentations for them.

**Mr. Abdul Razaque** is PhD student of computer science and Engineering department in University of Bridgeport. His current research interests include the design and development of learning environment to support the learning about heterogamous domain, collaborative discovery learning and the development of mobile applications to support mobile collaborative learning (MCL), congestion mechanism of transmission of control protocol including various existing variants, delivery of multimedia applications. He has published over 45 research contributions in refereed conferences, international journals and books.

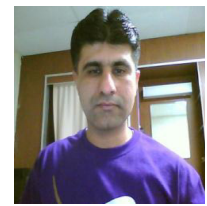

He has also presented his work more than 20 countries. During the last two years.
He has been working as a program committee member in IEEE, IET, ICCAIE, ICOS, ISIEA and Mosharka International conference. Abdul Razaque is member of the IEEE, ACM and Springer Abdul Razaque served as Assistant Professor at federal Directorate of Education, Islamabad. He completed his Bachelor and Master degree in computer science from university of Sindh in 2002. He obtained another Master degree with specialization of multimedia and communication (MC) from Mohammed Ali Jinnah University, Pakistan in 2008. Abdul Razaque has been directly involved in design and development of mobile applications to support learning environments to meet pedagogical needs of schools, colleges, universities and various organizations.

**Dr. Khaled Elleithy**: is the Associate Dean for Graduate Studies in the School of Engineering at the University of Bridgeport. His research interests are in the areas of, network security, mobile wireless communications formal approaches for design and verification and Mobile collaborative learning. He has published more than one two hundreds research papers in international journals and conferences in his areas of expertise.

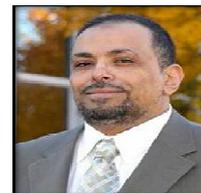

Dr. Elleithy is the co-chair of International Joint Conferences on Computer, Information, and Systems Sciences, and Engineering (CISSE).CISSE is the first Engineering/ Computing and Systems Research E-Conference in the world to be completely conducted online in real-time via the internet and was successfully running for four years. Dr. Elleithy is the editor or co-editor of 10 books published by Springer for advances on Innovations and Advanced Techniques in Systems, Computing Sciences and Software.

Dr. Elleithy received the B.Sc. degree in computer science and automatic control from Alexandria University in 1983, the MS Degree in computer networks from the same university in 1986, and the MS and Ph.D. degrees in computer science from The Center for Advanced Computer Studies in the University of Louisiana at Lafayette in 1988 and 1990, respectively. He received the award of "Distinguished Professor of the Year", University of Bridgeport, during the academic year 2006-2007.